\documentclass[preprint,showpacs,preprintnumbers,amsmath,amssymb]{revtex4}

\usepackage{graphicx}
\usepackage{dcolumn}
\usepackage{bm}

\begin{document}

\preprint{APS/123-QED}

\title{Numerical Study of Energy Loss by a Nanomechanical Oscillator Coupled to a Cooper Pair Box}


\author{Rakesh P. Tiwari and D. Stroud}
\affiliation{%
Department of Physics, Ohio State University, Columbus, OH 43210\\
}%

\date{\today}

\begin{abstract}

We calculate the dynamics of a nanomechanical oscillator (NMO) coupled capacitively to a Cooper pair box (CPB), by solving a stochastic Schrodinger equation with two Lindblad operators.  Both the NMO and the CPB are assumed dissipative,
and the coupling is treated within the rotating wave approximation.
We show numerically that, if the CPB decay time is smaller than the NMO decay time, the coupled NMO will lose energy faster, and the coupled CPB more slowly, than do the uncoupled NMO and CPB.  The results show that the efficiency of energy loss by an NMO
can be substantially increased if the NMO is coupled to a CPB.  

\end{abstract}

\maketitle

\section{Introduction}

Observing quantum effects in a mechanical object has been a long-sought goal, because
such effects would represent a macroscopic manifestation of quantum mechanics\cite{phystoday}.  Nanomechanical oscillators (NMO's) with frequencies $\omega/(2\pi)$
as high as 1 GHz have recently been realized in the laboratory\cite{huang}.   It is possible to observe quantum
effects in such objects, provided that they are cooled to 
temperatures $T$ such that $k_BT$ is near or slightly above $\hbar\omega$\cite{katz}.
For a realistic NMO frequency of 500 MHz, this 
corresponds to a $T$ of a few mK.

In this paper, we numerically analyze energy loss by a NMO capacitively coupled to a dissipative 
Cooper pair box (CPB).  A CPB consists of a small Josephson junction whose Josephson
energy is much smaller than its charging energy, and which
is voltage-biased so that only the two lowest energy levels are experimentally accessible.   Naik {\it et al.}\cite{clerk} have recently suggested
that an NMO could be cooled by capacitively coupling it 
to a CPB.  In the present work, we numerically demonstrate an increased rate of energy loss by this mechanism, by solving a time-dependent Schr\"{o}dinger equation with a stochastic term.  
We also relate
the rate of energy loss to the coupling constant and the NMO and CPB damping rates.  
Several studies of the coupled NMO/CPB system have already been carried out\cite{irish,armour,marquardt},
as have studies of the formally similar system consisting of a 
CPB couplied to a single-mode electromagnetic cavity\cite{blais}.  If both the CPB and the cavity are
dissipative, this system is closely analogous to the one studied here\cite{blais}.  

\section{Formalism}

Dissipation in quantum systems was treated by Caldeira and Leggett, who studied the suppression of quantum tunneling out of a metastable state by dissipation\cite{cl}.  Later, Leggett {\it et al.}
studied the dynamics of a dissipative TLS \cite{leggettetal}.   In both of these approaches, 
the dissipation is treated as arising from coupling to
a bath of harmonic oscillators.  An alternative method starts from 
Lindblad's quantum-mechanical master equation for the density operator of an open system\cite{lindblad}. 
An equivalent approach is to start from
a pure-state Schr\"{o}dinger equation, which contains a stochastic term to represent
the effect of dissipation\cite{diosi}.  Here, we use this approach
to treat energy loss by an NMO coupled to a CPB.

We assume that the Hamiltonian for the coupled NMO/CPB system is given by
\begin{eqnarray}
H&=&H_{CPB}+H_{NMO}+H_{I}+H_{\kappa}+H_{\gamma},\\
H_{CPB}&=&4E_{C}(n_g-\frac{1}{2})\sigma_{z}-\frac{1}{2}E_J\sigma_x,\\
H_{NMO}&=&\hbar\omega_{0}({a}^{\dagger}{a}+\frac{1}{2}),\\
H_{I}&=&\hbar g (a^{\dagger}+{a}){\sigma}_z. \label{eq:ham}
\end{eqnarray}
Here ${\sigma}_z$ and ${\sigma}_x$ are the Pauli spin matrices, which act on the two charge states of the CPB; 
$n_g=(C_bV_b+C_gV_g)/2e$, where $C_b$ and $V_b$ are the bias capacitance and voltage of the CPB;
$C_g$ and $V_g$ are the capacitance and voltage between the NMO and the CPB; $E_C$ and $E_J$ are the Coulomb and Josephson energies for the CPB; $\omega_0$ is the frequency of the fundamental
mode of the NMO; ${a}$ and ${a}^{\dagger}$ are the usual annihilation and creation operator for the fundamental mode of the NMO, and $g=-4E_C(C_gV_g)\Delta x_{zp}/(2e\hbar d)$ represents
the coupling constant for capacitive coupling between the NMO and the CPB.  The zero-point displacement of the NMO is given by
$\Delta x_{zp}=\sqrt{\hbar/2m\omega_0}$, where $m$ is the effective mass of the NMO and $d$ is the distance between the NMO and the CPB\cite{irish2}. 
The last two terms in $H$ represent the effects of dissipation.  $H_{\kappa}$ represents the dissipation in the NMO,
which is assumed to be characterized by a decay rate $\kappa=\omega_0/Q$, where $Q$ is the quality factor of the NMO. $H_{\gamma}$ represents the dissipation in the CPB, which is described by a decay rate $\gamma$. 
We also assume that only the fundamental mode of the NMO is coupled to CPB.  The precise forms of
$H_\kappa$ and $H_{\gamma}$ are given below. 

The Hamiltonian (1), in the absence of dissipation, has
been discussed, e.\ g., in Refs.\ \cite{irish} (which shows a sketch of a typical experimental
configuration), \cite{armour} and \cite{makhlin}.
We assume that the CPB is biased to $n_g=1/2$, 
where it would be degenerate in the absence of 
Josephson coupling, and work in the coordinate system obtained from
that of eq.\ (\ref{eq:ham}) by a rotation of $\pi/2$ about y-axis. In this coordinate system $H_{CPB}=(E_{J}/2){\sigma}_z$ and $H_{I}=\hbar g({a}+{a}^{\dagger}){\sigma}_x$.  
Moreover, in this coordinate system, the two eigenstates of $H_{CPB}$ represent symmetric and antisymmetric
superpositions of charge states.  In addition, 
we will make the rotating wave approximation (RWA) to obtain the Hamiltonian used in all our calculations. This Hamiltonian is given by
\begin{equation}
H=\frac{\hbar\Omega}{2}{\sigma}_z +\hbar\omega_0\left({a}^{\dagger}{a}+\frac{1}{2}\right)+
\hbar g({a}^{\dagger}{\sigma}^{-}+{\sigma}^{+}{a}) +H_{\kappa}+H_{\gamma},
\label{eq:rwa}
\end{equation}
where $\Omega=E_J/\hbar$ and $\sigma^{\pm} = (\sigma_x \pm i\sigma_y)/2$.  
Thus, we neglect  the counterrotating terms 
$\hbar g({a}^\dagger{\sigma}^+ + {a}{\sigma}^-)$ 
If $\Omega \sim \omega$ and the coupling is sufficiently weak, the RWA is believed justified \cite{irish2}. 
For this reason, we believe that the RWA is also reasonable
when the two damping terms are included.  

We include dissipation in both the CPB and the NMO by using a stochastic time-dependent
Schr\"{o}dinger equation.  This approach originates in the von Neumann equation for the total density matrix $\rho$
of any system: $\dot{\rho}=-\frac{i}{\hbar}[H,\rho]$.
The Hamiltonian of eq.\ (\ref{eq:rwa}) consists of two parts: the CPB plus NMO, described by the first three terms, and the dissipative parts, described by $H_{\kappa}$ and $H_{\gamma}$, which represent the damping. If the evolution of $\rho$ is considered to be Markovian, 
and if the coupling between the open system and the baths is weak, 
the reduced density matrix $\rho_{sys}$ of the CPB plus NMO can be shown to be
described by the following master equation\cite{wm}:
\begin{equation}
\dot{\rho}_{sys}=-\frac{i}{\hbar}[H-H_{\kappa}-H_{\gamma},\rho_{sys}]-\frac{1}{2}\sum_{m=\{\kappa,\gamma\}}(L_m^{\dagger}L_m\rho_{sys}+\rho_{sys}L_m^{\dagger}L_{m}-2L_m\rho 
L_m^{\dagger}).
\label{eq:master}
\end{equation}
Here the $L_m$ are so-called Lindblad operators, which are in general non-Hermitian and which describe dissipation in CPB plus NMO system. In the present work, we include only the two Lindblad operators $L_\kappa=\sqrt{\kappa}{a}$ and $L_\gamma=\sqrt{\gamma}{\sigma}^{-}$, which represent
dissipation within the NMO and the CPB.  The inclusion of only these two Lindblad operators
is appropriate if $T$ is sufficiently low 
(typically $k_B T \ll \hbar\omega_0$), 
and if other dissipative processes can be neglected.   

Eq.\ (\ref{eq:master}) is a matrix equation for the density operator $\rho_{sys}$.  If we  truncate the state
space by including only the lowest $N$ vibrational states, then $\rho_{sys}$ is a 
square matrix of size $2N \times 2N$.  Eq.\ (\ref{eq:master}) then leads to (2N)$^2$ coupled differential equations for
the elements of $\rho_{sys}$.   An alternative computational scheme is to use a stochastic pure state representation \cite{diosi,gp} for the master 
equation. In this representation, eq.\ (\ref{eq:master}) reduces to a stochastic equation of motion for the state vector 
$\left|\psi\right\rangle$ given by  \cite{blais,schack}
\begin{eqnarray} 
\left|d\psi\right\rangle=&-&\frac{i}{\hbar}\left(H-H_\kappa-H_\gamma\right)\left|\psi\right\rangle dt + \sum_{m}\left(L_m-\left\langle L_m\right\rangle_\psi\right)\left|\psi\right\rangle d\xi_m \nonumber \\
&-&\frac{1}{2}\sum_m\left(L_m^{\dagger}L_m+\left\langle L_m^\dagger\right\rangle_\psi\left\langle L_m\right\rangle_\psi-2\left\langle L_m^\dagger\right\rangle_\psi L_m\right)\left|\psi\right\rangle dt, 
\label{eq:4}
\end{eqnarray}
where $\left\langle L_m\right\rangle_\psi$ is the expectation value of $L_m$ in state $\left|\psi\right\rangle$,
and $|d\psi\rangle$ is the change in $|\psi\rangle$ in time dt. 
In eq.\ (\ref{eq:4}), the first sum represents random fluctuations due to the interaction of the system with the baths, while the second denotes the (non-random) \textit{drift} of the state vector due to those baths. The $d\xi_m$ are independent complex differential random variables representing a complex normalized Wiener process, whose
ensemble averages satisfy
\begin{eqnarray}
\overline{d\xi_m}=\overline{d\xi_m d\xi_n}=0, \\
\overline{d\xi_m^{\ast}d\xi_n}=\delta_{mn}dt,
\label{eq:5}
\end{eqnarray}
where the overbar denotes an ensemble average over realizations of the random variables $d\xi_n$.

Eqs.\ (\ref{eq:4}) represent only $2N$ coupled differential equations for the components of $\left|\psi\right\rangle$, rather than $(2N)^2$ as in the density matrix formulation.  The price paid is that the
differential equations are stochastic, and thus must be averaged over many realizations.  But 
such stochastic equations can be solved very efficiently, even including this averaging.  Thus, we
use this stochastic  approach in the following.

To see how this procedure can lead to faster energy loss by the NMO, suppose that the initial 
state vector is $|\psi\rangle
= |\alpha, n\rangle$, where $\alpha (= 0$ or $1$) indicates an eigenstate of $H_{CPB}$ and $n$ represents
the initial number of excitations in the NMO. 
Thus, the NMO is assumed to be initially in a Fock (or number) state.
As time progresses the NMO and CPB become entangled, and the wave function $|\psi\rangle$
spreads out over an increasing part of the accessible 2N-dimensional Hilbert space.  
After some time $t$, $\left|\psi(t)\right\rangle$ is a superposition of many states $|\alpha, m\rangle$, with
$m \leq n+1$; the weight of each component is determined by eq.\ (\ref{eq:4}).  Several
experimental methods for preparing an NMO in a specified Fock state have been proposed \cite{irish,santamore}.

\section{Numerical Results}

We have solved eq.\ (\ref{eq:4}) at resonance for different values of $\kappa$ and $\gamma$, using the
value $g=0.1\omega_0$, by straightforward Euler integration, using a time step $dt = 5\times 10^{-5}$ns$= 2.5\times 10^{-5}(2\pi/\omega_0)$.
From these solutions, we compute various averages of interest, such as $\overline{\langle{\sigma}_z(t)\rangle}$ and $\overline{\langle{a}^{\dagger}{a}(t)\rangle}$, as functions of time $t$.
The triangular brackets and overbars denote quantum-mechanical and noise averages, respectively.  For
conciseness, we denote these quantities simply $\overline{\sigma_z}(t)$ and $\overline{n}(t)$. 
The results are averaged over 100 realizations of the noise, which appear sufficient to give reasonably
smooth results.
We carry out calculations assuming $\Omega/(2\pi) = \omega_0/(2\pi) = 500$ MHz,
$g/(2\pi) = -50$ MHz, $\kappa = 0.0005$ (ns)$^{-1}$and $\gamma = 0.05$ (ns)$^{-1}$.  These values of $\kappa$ and $\gamma$ correspond
to decay times of the uncoupled NMO and CPB of 2000 ns (corresponding to a $Q$ factor of $10^3$) and 20 ns, respectively, as used in Ref.\ \cite{nakamura}.  
A much larger decay time for the CPB has been achieved in some more recent experiments
(see, e.\ g., Wallraff {\it et al.}\cite{wallraff}, who find an energy relaxation rate of around
7$\mu$sec).  However, one does not want such a long decay time for the CPB, since this will reduce
the rate at which the NMO loses energy when coupled to the CPB.  We have, therefore, carried out
our calculations using the original parameters of Ref.\ \cite{nakamura}.  

  
We first carried out calculations of the uncoupled CPB and
NMO, using the stochastic Schr\"{o}dinger equation, starting from the state  
$|1,20\rangle$.  In this case, the CPB and NMO each decay exponentially
with decay rates 20 (ns)$^{-1}$ and 2000 (ns)$^{-1}$, 
respectively, as shown in Figs.\ 1 and 2.  Next, we turned on the coupling $g$, so that
$g/(2\pi) = -50$ MHz, with results shown in Figs.\ 1 and 2 for $\overline{\sigma_z}(t)$ and $\overline{n}(t)$. 

\section{Discussion}

We now discuss the results.  In both cases, we use $\gamma \gg \kappa$ of the NMO, as in experiment\cite{nakamura,irish}.  
Fig.\ 1 shows that, in the presence of coupling, the decay rate of the CPB is substantially reduced, relative to the uncoupled CPB.  The reduction in the decay rate is greatest when the damping in the NMO is smallest.
Besides this reduction in decay rate, Fig.\ 1 shows the expected Rabi oscillations in $\overline{\sigma_z}$ for all
three calculations involving nonzero $g$.  These Rabi oscillations, of
frequency $\omega_R \propto \sqrt{n+1}g$\cite{eberly}, occur between the states $|1,n\rangle$
and $|0, n+1\rangle$.   They are the well-known results of solving the
time-dependent Schr\"{o}dinger equation for the present Hamiltonian in the {\em absence} of dissipation; our
calculations show that they still occur with dissipation present.    

By contrast, Fig.\ 2 shows that, when the coupling is turned on, the decay rate of the NMO is {\it increased}
relative to that of the uncoupled NMO.  Furthermore, the decay rate is the largest when the damping rate of
the CPB is largest.  Once again, we see Rabi oscillations in $\overline{n}(t)$. 

The results of both Figs.\ 1 and 2 can be understood qualitatively when $\gamma \gg \kappa$.
The coupling allows energy to be transferred periodically between the CPB and the NMO.
Since the NMO has a much lower decay rate than that of the CPB, this new channel
should reduce the effective decay rate of the CPB, as we observed numerically.  
Similarly, this transfer should {\em increase} the effective damping rate of the NMO, again as seen
numerically.  Furthermore as the damping rate $\gamma$ of the CPB
increases, the effective damping rate of the NMO should also increase, again as seen
numerically.

Finally, in Fig.\ 3, we show how this behavior depends on the assumed initial Fock state of the NMO.  In
the three cases shown, we start from state $|1, n\rangle$ with four different initial Fock states, 
$n(0)$, of the NMO.   
$\overline{n}(t)/n(0)$ decays roughly exponentially with time, with superimposed Rabi oscillations.  The exponential decay rate is consideably smaller for
$n(0)=1$ than for the other cases.  This difference occurs, we believe, because, in the absence of damping, the  coupling between the CPB and NMO is proportional $\hbar g\sqrt{n+1}$ (this is the splitting between states
$|1, n\rangle$ and $|0, n+1\rangle$ in the RWA in the absence of damping).  


We now discuss the possible relation between the present results and
energy loss by a real NMO.  
We have considered
the time-dependent NMO/CPB system, starting from the state $|1,n\rangle$.  Our approach permits us to calculate 
the time needed for the NMO, starting from this excited state, to approach its equilibrium temperature, given that the intrinsic damping within the NMO is very small.
   If the NMO is coupled to a much more heavily damped CPB, 
we find that the equilibration process is greatly speeded up.  
This result might be useful in designing ways of rapidly equilibrating an NMO, initially in an
excited state, to a low ambient temperature.

We also find that the rate of equilibration 
of the
CPB is {\it reduced} when it is coupled to an NMO.  Our results are consistent with those of Trees {\it et al.}\cite{btrees}, who used a very different approach.  These workers studied a current-biased Josephson junction capacitively coupled to an oscillator, in the presence of dissipation, using the formalism of Caldeira and Leggett\cite{cl}, and found that the CPB is less damped when it is coupled to an NMO.   Our results show the same behavior, using a formally quite different stochastic differential equation.  We also find an additional result, not shown
in Ref.\ \cite{btrees}: for a given coupling constant, the CPB is damped
most slowly when the damping of the NMO is smallest (see Fig.\ 1).

It should also be pointed out that we have included only two Lindblad operators in our calculation, namely,
those involving the constants $\kappa$ and $\gamma$.  Other Lindblad operators could readily be included, such as
ones describing finite temperature and pure dephasing\cite{blais}.  In the present calculation, $\kappa$ and $\gamma$ parametrize energy loss from the NMO and the CPB to their thermal environments.  Another point is that the regime
we have considered, in which the CPB is more heavily damped than the NMO, is not necessarily the regime in which
a CPB is typically studied.  When one wishes to use the CPB as an element in a quantum computing geometry,
it is usually desirable to have a CPB with as little dissipation, and as long a decoherence time, as possible.  In the present work, by contrast, we are
envisioning a setup where the CPB is being used as a means of rapidly extracting energy from a high-Q NMO, so as to observe the expected quantum effects in the NMO.  As shown by our calculations, this requires a CPB which is more heavily damped than the NMO.  Finally, we should note that our calculation is applicable, in principle, to energy loss by any harmonic oscillator mode coupled to any two-level system, provided that the model Hamiltonian describing the
interactions and energy losses is that used in the present work.

Next, we briefly comment on the initial conditions used in these calculations.  We have assumed that the NMO begins in a Fock state.  In some cases, a more appropriate 
initial condition might be one in which the initial state of the oscillator is a mixture of Fock
states weighted according to a suitable temperature. 
If this calculation were carried out, it might represent a study of the rate of cooling of an NMO, initially at some temperature $T$, if it were placed in a thermal bath at a much lower temperature and also coupled to an NMO at a similarly lower temperature.   Another possible initial state might be a coherent state of the oscillator, i.\ e., an eigenstate of $a$.  Both initial states may be achievable experimentally.  Calculations starting from such mixed states would be more demanding numerically, using the present approach, but will probably lead to qualitatively similar results 
regarding the rate of energy loss of both the NMO and the CPB. 

Finally, we comment on our use of the RWA.  The RWA is believed
justified when the coupling between the NMO and CPB is reasonably small, compared to $\hbar\omega_0$, and also
provided the NMO and CPB are in resonance.  Both conditions are satisfied in the present calculation.  We have carried out similar RWA calculations
off resonance, with results similar to those shown in Figs.\ 1-3, but we
believe that the RWA approximation is less accurate in this case.  To go beyond the RWA leads to a considerably more complicated set of stochastic differential equations.
To confirm our results, we have, in fact, gone beyond the RWA for the parameters used in the calculations discussed above (NMO and CPB in resonance, relatively weak coupling), solving eqs.\ (7) including all the counterrotating terms. In this case, it proves more convenient to solve numerically for $|\psi^\prime\rangle = \exp[-i(H - H_\kappa - H_\gamma)t/\hbar]|\psi\rangle$, where $H$ is given by eq.\ (5), then transform back to get $|\psi\rangle$.  The resulting time dependent averages, for these parameters, are nearly indistinguishable from those shown in Figs.\ 1-3, showing that the RWA is indeed an excellent
approximation, as expected, for this case.  For substantially stronger coupling between the NMO and the CPB, or
for frequencies far off resonance, we have not yet succeeded in obtaining a converged solution for $|\psi(t)\rangle$
when the counterrotating terms are included, showing that, once again as expected, the RWA is inaccurate in this
regime.

To summarize, we have demonstrated, within the RWA, and using a suitable time-dependent Schr\"{o}dinger 
equation with stochastic terms, that the rate of equilibration of an NMO with a thermal bath can be considerably 
speeded up by coupling the NMO to a CPB.  Using the same approach,
we have demonstrated that the decay rate of the CPB can be easily reduced by coupling it to a high $Q$ factor cavity,
such as a nanomechanical oscillator; this latter result is consistent with previous calculations\cite{btrees} using
a very different approach.  Both of these effects may be very useful experimentally.  In particular, the use of a 
CPB as a means of removing energy from an NMO may be helpful in observing macroscopic quantum effects in such
oscillators.

This work has been supported by the NSF, through grant DMR-04-13395, and also benefited from the facilities of the Ohio Supercomputer Center.  We thank Professor B.\ R.\ Trees for valuable
conversations.

\begin{figure}[h]
\begin{center}
\includegraphics[scale=0.5,angle=270]{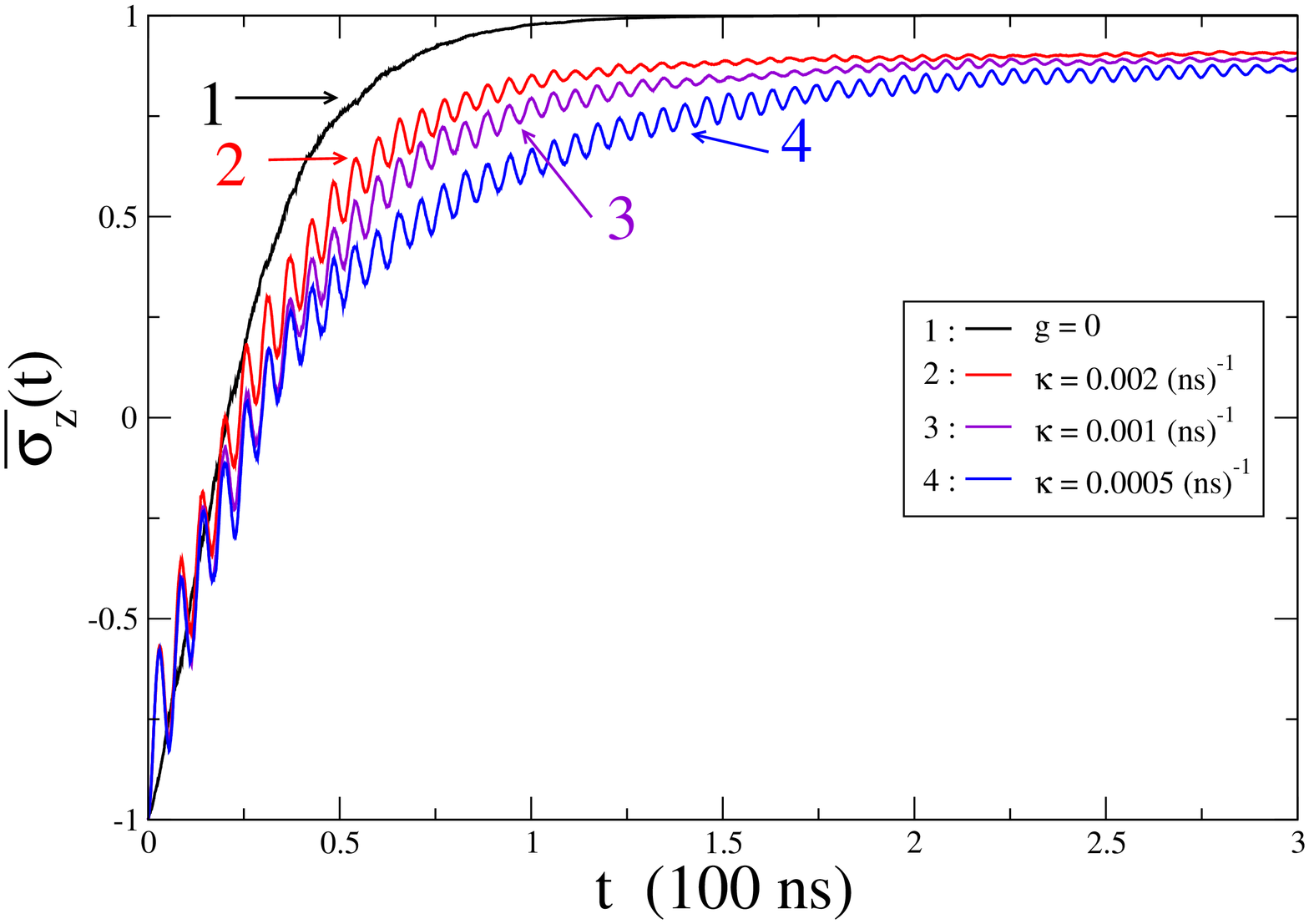}
\caption{(Color online)  Calculated $\overline{\sigma_z}(t)$ versus time $t$, averaged over 100
realizations of the noise, for the case of zero coupling to the NMO
(g = 0), and for fixed $g/(2\pi)$ = -(50 MHz), and
three different values of the NMO damping parameter $\kappa$. } \label{fig:1}
\end{center}

\end{figure}
\begin{figure}[h]
\begin{center}
\includegraphics[scale=0.5,angle=270]{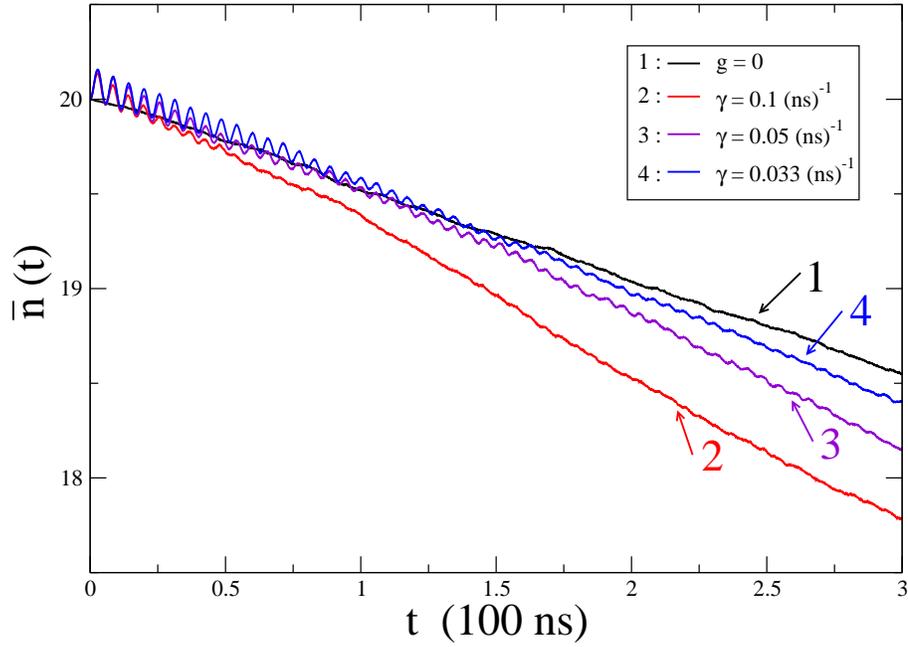}
\caption{(Color online)  Expected number of excitations in the NMO, $\overline{n}(t)$, plotted versus time.  
$\overline{n}(t)$ is averaged over 100 noise realizations, and is shown for zero coupling (g = 0) 
and for $g/(2\pi)$ = - 50 MHz and three different values of the
CPB damping parameter $\gamma$ as indicated.} \label{fig:2}
\end{center}
\end{figure}

\begin{figure}[h]
\begin{center}
\includegraphics[scale=0.5,angle=270]{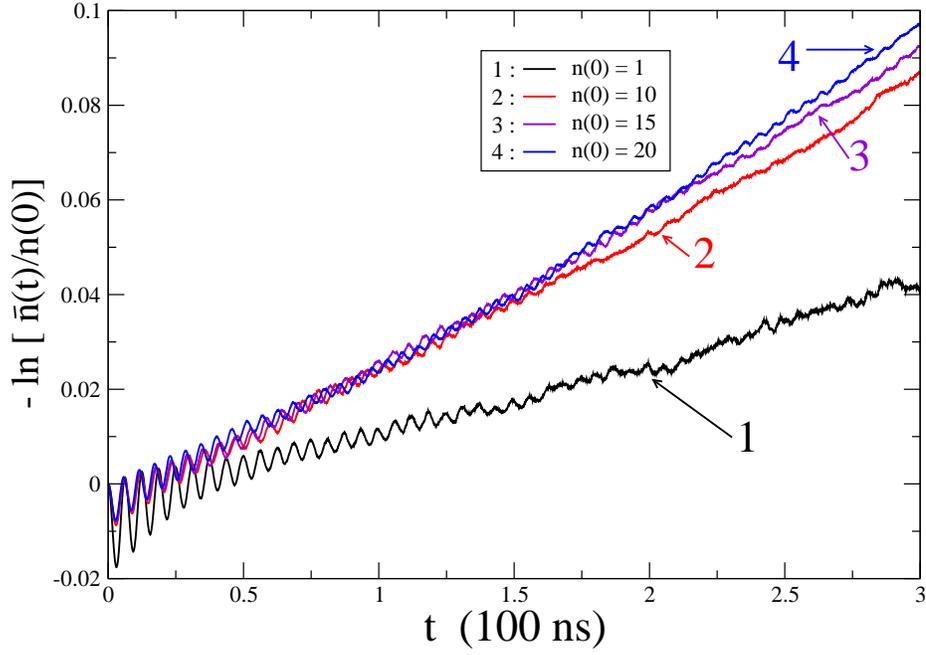}
\caption{(Color online)  $-\ln[\overline{n}(t)/n(0)]$ for different initial Fock states of the NMO.  In
all cases, $g/(2\pi)$ = -50MHz, $\gamma = 0.05 (ns)^{-1}$ and $\kappa = 0.0005 (ns)^{-1}$.} \label{fig:3}
\end{center}
\end{figure}

\end{document}